\documentclass[aps,prl,twocolumn,groupedaddress,showpacs,amssymb,superscriptaddress]{revtex4}
 
\usepackage{graphicx}
\usepackage{epsfig}

\begin{document}
\bibliographystyle{apsrev}
 
\title{Switching the sign of Josephson current through Aharonov-Bohm interferometry}

\author{Fabrizio~Dolcini}  
\email{f.dolcini@sns.it}
 
\author{Francesco Giazotto}

\email{f.giazotto@sns.it}
\affiliation{Scuola Normale Superiore and   NEST CNR-INFM, I-56126 Pisa, Italy}

\date{\today}

\begin{abstract}
We investigate the DC Josephson effect in a superconductor-normal metal-superconductor junction where the normal region consists of a ballistic  ring. We show that a fully controllable $\pi$-junction can be realized through the electro-magnetostatic Aharonov-Bohm effect in the ring. The sign and the magnitude of the supercurrent can be tuned by varying the magnetic flux and the gate voltage applied to one   arm, around suitable values. The implementation  in a realistic set-up is discussed.

\end{abstract}
 
\pacs{73.23.-b, 74.50.+r, 85.25.Cp, 85.25.-j}

\maketitle

The broad interest in mesoscopic physics has recently spurred a renewed impulse in the context of Josephson effect. Because of its large spectrum of applications to nanotechnology, the art of manipulating the supercurrent is presently under the spotlight~\cite{golubov}. Josephson field-effect transistors, for instance, have been   proposed~\cite{clark} and   realized   both with semiconductors and carbon nanotubes~\cite{Jofet}. A growing interest is nowadays devoted to the issue of supercurrent sign reversal: a $\pi$-junctions state, i.e., a Josephson current flowing in the direction opposite to the phase difference between the superconductors, has already been obtained with ferromagnet-superconductor (FS) junctions~\cite{FS}. In these systems  the sign of the current flow depends on the F-layer thickness, which cannot be varied during an experiment. In view of large-scale technological applications, the tunability of a system plays instead a crucial role, and the realization of \emph{controllable} $\pi$-junctions represents a major challenge. To this end, two approaches have been explored so far: tunable $\pi$-states have been realized either by driving far from equilibrium the junction quasi-particle  distribution function through current injection from additional terminals~\cite{noneq}, or by exploiting the   electron-electron correlation in a quantum dot, where the electron number can be tuned through a gate~\cite{defrance}.

\begin{figure}
\includegraphics[width=\columnwidth,clip]{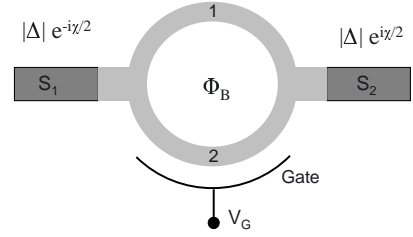}
\caption{\label{AB-setup} Scheme of the Aharonov-Bohm Josephson junction. Light gray region indicates the normal metal ring, whereas dark gray denotes the superconducting leads.}
\end{figure}

In this letter, we present a different operational principle to implement a controllable $\pi$-junction, which is based on the electro-magnetostatic Aharonov-Bohm (AB) effect in a ballistic normal (N)  ring coupled to two superconducting (S) leads. The proposed set-up is shown in Fig.~\ref{AB-setup}: the ring is threated by a   magnetic flux~$\Phi_B$, and a gate voltage~$V_G$ is applied to one of the  arms. Exploiting the magnetic or the electrostatic AB effect, one can tune the transmission of the ring and therefore manipulate the magnitude of the Josephson current. Here we show that, if {\it both} the magnetic and the electric field are suitably applied,  also the {\it sign}  of  the supercurrent can be changed. This leads to a fully controllable  system  operating at equilibrium, which exhibits  potential for the implementation  of low-dissipation transistors,  quantum interference devices,  as well as reduced-noise qubits~\cite{qc}. Ideal candidates for the realization of the proposed set-up are intermediate/long SNS junctions fabricated with In$_x$Ga$_{1-x}$ (with $x\geq 0.75$)~\cite{carillo} or InAs~\cite{kasper} rings, since they allow ballistic electron transport and lack of a Schottky barrier when contacted to metals,~e.g. superconducting~Nb. \\
We describe the system by setting the superconducting order parameter as $\Delta(x)=|\Delta| e^{-i \chi/2}$ in the left lead, $\Delta(x)=|\Delta| e^{+i \chi/2}$ in the right lead, and $\Delta(x)=0$ in the N region, where~$\chi$ is the phase difference between the S electrodes.
The Josephson current $J(\chi)$ of an SNS junction with a N-region characterized by a Fermi velocity $v_F^N$ and a length~$L$ can be written~\cite{BeenPRL91,bardeen} as a sum of two contributions $J_{\rm d}(\chi)$ and $J_{\rm c}(\chi)$,  arising from the discrete Andreev levels and the continuous spectrum  respectively. Explicitly, $J_{\rm d,c}(\chi)= (2 e E_L/\hbar)  j_{\rm d,c}(\chi)$, where $E_L= \hbar v^N_F/L$ is the energy associated with $L$ (here the length of each ring arm),   and
\begin{eqnarray}
j_{\rm d}(\chi) &=& \sum_{p} \tanh(\frac{\varepsilon_p \beta_L}{2}) \left( -\frac{\partial \varepsilon_p}{\partial \chi} \right) \label{curr-Been-disc} \\
j_{\rm c}(\chi) &=& \frac{2}{\pi \beta_L} \int_{|\Delta_L|}^{\infty} \hspace{-.2cm} d\varepsilon \ln [ 2 \cosh(\frac{\varepsilon \beta_L}{2})] \frac{\partial {\rm Im} \left[ H(\varepsilon; \chi) \right]}{\partial \varepsilon} .  \label{curr-Been-cont}
\end{eqnarray}
In Eqs.~(\ref{curr-Been-disc})-(\ref{curr-Been-cont})   energies are expressed in terms of $E_L$, so that  $|\Delta_L|=|\Delta|/E_L$, $\varepsilon=(E-E_F)/E_L$ is the energy variation with respect to the Fermi level $E_F$ of leads and N-region, and $\beta_L=E_L/k_B T$ is  the inverse temperature. The Andreev levels  $\varepsilon_p(\chi) \in [0; |\Delta_L|]$ in (\ref{curr-Been-disc}) are  determined by the solutions of $D(\varepsilon;\chi)=0$, where  the function
\begin{equation}
D(\varepsilon;\chi)=\det [\openone-A(\varepsilon) r_A^{*}(\chi) \mathsf{S}(\varepsilon) r_A(\chi) \mathsf{S}^{*}(-\varepsilon) ] \label{Ddef}
\end{equation}
accounts both for the dynamics in the N-region, through the particle (hole) scattering matrix $\mathsf{S}(\varepsilon)$ ($\mathsf{S}^{*}(-\varepsilon)$), and for  the Andreev reflections at the S contacts through the matrix $r_A(\chi)={\rm diag}( e^{i \chi/2} , e^{-i \chi/2} )$ and the amplitude $A(\varepsilon)$, which equals $\exp(-2 i \arccos(\varepsilon/|\Delta_L|))$ for $|\varepsilon| < |\Delta_L|$  and $(\varepsilon-\sqrt{\varepsilon^2-|\Delta_L|^2})/(\varepsilon+\sqrt{\varepsilon^2-|\Delta_L|^2})$ for $|\varepsilon| > |\Delta_L|$. Finally, the function $H$ appearing in (\ref{curr-Been-cont}) reads
\begin{equation}
H(\varepsilon; \chi) = {\partial_\chi} \ln D(\varepsilon;\chi). \label{Hdef}
\end{equation}

Although general, Eqs.~(\ref{curr-Been-disc})-(\ref{curr-Been-cont})  do not allow a straightforward evaluation of $J(\chi)$ for an arbitrary $\mathsf{S}$-matrix, for the determination of the Andreev levels  may be a formidable task. In the limit of short junction $|\Delta_L| \ll 1$, however, Beenakker showed that the contribution (\ref{curr-Been-cont}) vanishes, and that the evaluation of the Andreev levels in (\ref{curr-Been-disc}) is considerably simplified. He thus proved~\cite{BeenPRL91} that in this limit and for a symmetric $\mathsf{S}$-matrix,  the Josephson current acquires a simple expression in terms of the transmission coefficients $\mathcal{T}_i$'s of  $\mathsf{S}$. 
In a typical experimental realization of the set-up of Fig.~\ref{AB-setup}, however,  the  short junction regime $|\Delta_L| \ll 1$ is not achieved; as observed above, one rather has $|\Delta_L|  \gtrsim 1$. Furthermore, in the presence of a magnetic flux, the ring $\mathsf{S}$-matrix is not symmetric in general. It is therefore important to analyze the behavior of the Josephson current also beyond the  short-junction limit, and without assuming the symmetry of the $\mathsf{S}$-matrix. To this purpose, it is suitable to rewrite  Eqs.~(\ref{curr-Been-disc})-(\ref{curr-Been-cont}) in a different way. Here we briefly sketch the main steps of our strategy, based on  analytic continuation of Eq.~(\ref{Hdef}) in the complex plane. We first observe that, due to causality~\cite{roman}, the scattering matrix $\mathsf{S}(\varepsilon)$  has an analytic continuation $\mathsf{S}(z)$ in the upper complex half-plane ${\rm Im}(z) > 0$. Then, elementary properties of holomorphic functions lead to prove that  $H_{\rm R}(z;\chi)={\partial_\chi} \ln D_{\rm R}(z;\chi)$ and $H_{\rm A}(z;\chi)= {\partial_\chi} \ln D_{\rm A}(z;\chi)$ with
\begin{eqnarray}
D_{\rm R}(z;\chi) =  \det \left[ \openone-A(z) \, r_A^{*}(\chi) \mathsf{S}(z) r_A(\chi) (\mathsf{S}(-z^{*}))^{*} \right] \label{Hret-def} \\  
D_{\rm A}(z;\chi) = \det \left[ \openone-A(z) \, r_A(\chi) (\mathsf{S}(z^*))^{*} r^{*}_A(\chi) \mathsf{S}(-z) \right]
\label{Hadv-def} \end{eqnarray}
are analytic continuations of (\ref{Hdef}) for ${\rm Im} (z) >0$ and ${\rm Im} (z) <0$  respectively. Here $A(z)$ is the continuation of $A(\varepsilon)$ with a branch-cut in  $z \in [ -|\Delta_L|; |\Delta_L| ]$.
The relation $H_{\rm R}(\varepsilon + i 0^{+})= H^{}_{\rm A}(\varepsilon + i 0^{-}) $ for $| \varepsilon | < |\Delta_L|$ stemming from (\ref{Hret-def})-(\ref{Hadv-def}), together with equality $\partial_\chi \varepsilon_p = - \partial_\chi D (\varepsilon_p; \chi)/ \partial_\varepsilon D (\varepsilon_p; \chi)$ for the Andreev levels, allows to rewrite Eq.~(\ref{curr-Been-disc}) as
\begin{eqnarray}
j_{\rm d}(\chi)= \frac{1}{2 \pi i} \sum_{\eta={\rm R,A}} \sum_p  \int_{\Gamma^p_\eta} \tanh(\frac{z\beta_L}{2}) H_{\rm \eta}(z,\chi) dz \, \, , \label{j-disc-new}
\end{eqnarray}
where $\Gamma^p_R$ ($\Gamma^p_A$) is a small semicircular contour around~$\varepsilon_p$, in the upper (lower) half-plane. Similarly, the relation $H_{\rm R}(\varepsilon + i 0^{+})= H^{*}_{\rm A}(\varepsilon + i 0^{-}) $ for $| \varepsilon | > |\Delta_L|$ stemming from (\ref{Hret-def})-(\ref{Hadv-def}) leads to cast Eq.~(\ref{curr-Been-cont})  in the same form as Eq.~(\ref{j-disc-new}), where the sum over $\Gamma^p_R$ ($\Gamma^p_A$) is replaced by a contour along the upper (lower) real axis range $\varepsilon \in [ |\Delta_L|; \infty]$. Finally, the analyticity of $H_{\rm R}(z;\chi)$ and $H_{\rm A}(z;\chi)$ allows to merge the above two contributions into 
\begin{equation}
J(\chi) = - \frac{4 e \, k_B T}{\hbar} \, \partial_\chi \sum_{m=0}^{\infty} \, {\rm Re} \ln  \left( D_{\rm R}(i w_m;\chi) \right) \, , \label{j-new}
\end{equation}
where $w_m=(2m+1) \pi/\beta_L$ are the Matsubara frequencies in units of $E_L/\hbar$. The relation $H_{\rm A}(-i w_n;\chi)=H^*_{\rm R}(i w_n;\chi)$ has been used. \\

Equation (\ref{j-new}), combined with Eq.~(\ref{Hret-def}), gives the Josephson current in terms of the N-region scattering matrix~$\mathsf{S}$.  Although equivalent to the original Eqs.~(\ref{curr-Been-disc})-(\ref{curr-Been-cont}), Eq.~(\ref{j-new}) does not require to determine the Andreev levels, yielding a major simplification in computing $J(\chi)$ for non-short junctions~\cite{limits}.  We shall now show the rich physical scenario arising from a non symmetric $\mathsf{S}$-matrix. In order to illustrate this, we focus here on the case of a single conduction channel, where the $\mathsf{S}$-matrix is a $2 \times 2$ unitary matrix. Any ${\mathsf{U}}(2)$ matrix can be univocaly  written as
\begin{equation}
\mathsf{S}(\varepsilon) = \xi(\varepsilon) \left( \begin{array}{ccc} \rho(\varepsilon) & & -i \tau(\varepsilon) \\ & & \\ -i \tau^{*}(\varepsilon) & & \rho^*(\varepsilon) \end{array} \right) \label{S2x2}
\end{equation}
\\where $|\rho(\varepsilon)|^2+|\tau(\varepsilon)|^2=1$ and $\xi$ is a radius (i.e., $|\xi|=1$).
Inserting Eq.~(\ref{S2x2}) into Eq.~(\ref{Ddef}) one obtains
\begin{eqnarray}
\lefteqn{D(\varepsilon;\chi) = 1+A^2(\varepsilon) e^{2 i \Psi(\varepsilon)} - 2 A(\varepsilon) e^{i \Psi(\varepsilon)} \times  } & & \label{Dret-fin} \\
&   &  \left[ {\rm Re} \left(\rho(\varepsilon) \rho^*(-\varepsilon)\right) + |\tau^{}(\varepsilon) \tau^{}(-\varepsilon)| \, \cos(\chi-\chi_0(\varepsilon)) \right] \nonumber ,
\end{eqnarray}
Notably, Eq.~(\ref{Dret-fin})  shows that the phase difference $\chi$ is renormalized by a shift $\chi_0$, related to the phase  of the amplitude $\tau(\varepsilon)= |\tau(\varepsilon)| \, \exp(i \phi_\tau(\varepsilon))$ through the relation
\begin{equation}
 \chi_0(\varepsilon)=\phi_\tau(\varepsilon)+\phi_\tau(-\varepsilon)    . \label{chi0def}
\end{equation}
From Eqs.~(\ref{chi0def}) it follows that $\chi_0$ is non vanishing only if $\tau$  is a complex number. At first sight, this property might seem a quite general feature of any quantum interferometer. This is, however, not the case, for the entries $\mathsf{S}_{ij}$ of the scattering matrix obey the micro-reversibility relation~\cite{roman} ${\mathsf{S}_{ij}}|_B={\mathsf{S}_{ji}}|_{-B}$, where $B$ is the magnetic field. Thus, for any  interferometer operating in the absence of magnetic field, the $\mathsf{S}$-matrix is symmetric, i.e., $\tau(\varepsilon)$ is real (see Eq.~(\ref{S2x2})). One thus has $\chi_0=0$ and the system always behaves as a $0$-junction.   By contrast, if $B \neq 0$, the time-reversal symmetry (TRS) is broken, opening the possibility to the asymmetry of the $\mathsf{S}$-matrix, and hence to a~$\chi_0 \neq 0$. Importantly, $B \neq 0$ is a necessary, but not sufficient condition to have a~$\pi$-state: in a symmetric ring threaded by a uniform $B$, for instance,   $\mathsf{S}$  is still symmetric and one has a 0-junction. If, however, the interferometer is suitably designed, an appropriate tuning of its parameters can lead to~$\chi_0 = \pi$,   inducing a switching to a $\pi$-state. \\

The electro-magnetostatic AB interferometer (see Fig.\ref{AB-setup}) is an illuminating example to describe  this effect. The explicit expression for the AB ring $\mathsf{S}$-matrix can be obtained with standard techniques~\cite{indiani-japuz} by combining the scattering matrices describing the $Y$-junctions with the propagation matrix along the ballistic arms. For simplicity we neglect here band curvature, fringe field effects, and spin-orbit interaction. Each contact is described by the $3 \times 3$ $Y$-junction matrix $\mathsf{S}^Y$~\cite{Y-Butt}, with entries $\mathsf{S}^Y_{11}=-\cos{\gamma}$, $\mathsf{S}^Y_{12}=\mathsf{S}^Y_{13}=\mathsf{S}^Y_{21}=\mathsf{S}^Y_{31}=\sin{\gamma}/\sqrt{2}$, $\mathsf{S}^Y_{22}=\mathsf{S}^Y_{33}=(\cos{\gamma}-1)/2$, and $\mathsf{S}^Y_{23}=\mathsf{S}^Y_{32}=(\cos{\gamma}+1)/2$, where the parameter $\gamma \in [0; \pi/2]$ accounts for the contact transmission, with $\gamma=\pi/2$ describing a fully transmitting contact and $\gamma \rightarrow 0$ the tunnel limit~\cite{Fazio}. The transmission of the two contacts will be assumed equal. The propagation along the two arms leads to the AB interferometry effect; right movers, for instance, acquire a phase $\exp(i(k L -\phi/2))$ along arm 1, and $\exp(i(k L +u+\phi/2))$ along arm 2, where $\phi=2 \pi e \Phi_B/h$ and $u=eV_G/E_L$ (see Fig.\ref{AB-setup}). Similarly for left-movers. After lengthy but standard algebra, one can compute the   $\mathsf{S}$-matrix (\ref{S2x2}) and the phase shift $\chi_0$, which turns out to fulfill
\begin{equation}
\cos{\chi_0(\varepsilon)}=\frac{{\rm Re} \, \tau_P(\varepsilon) {\rm Re} \, \tau_P(-\varepsilon)-{\rm Im} \, \tau_P(\varepsilon) {\rm Im} \, \tau_P(-\varepsilon)}{|\tau_P(\varepsilon) \tau_P(-\varepsilon)|} 
\label{coschi0}
\end{equation}
where $i \tau_P(\varepsilon)=\sum_{s=\pm} s C_s \exp[i s (\varepsilon+\kappa^\prime_F)]$, $C_{\pm}=\cos[(u\pm\phi)/2]$, and $\kappa^\prime_F=k_F L +u/2$, with $k_F$ the Fermi wave vector. Notice that~$\chi_0$ is independent of the contact transmission. For the purely electrostatic AB effect ($\phi=0$ and $u \neq 0$), we  obtain  $\chi_0=0$, as expected from the preservation of TRS. Moreover, Eq.~(\ref{coschi0}) yields $\chi_0=0$ also for the purely magnetic AB effect ($u=0$ and $\phi \neq 0$): this is due to the additional relation ${\mathsf{S}}_{12}|_B={\mathsf{S}}_{12}|_{-B}$, which holds in a symmetric ring  if $u=0$. Although this relation breaks down if the ring is realized asymmetrically, the arm length is not a tunable quantity. However, Eq.~(\ref{coschi0}) indicates that a much simpler way to achieve a controllable $\pi$-state is to combine the magnetic flux and the gate voltage: when both $\phi \neq 0$ and $u \neq 0$ a phase $\chi_0 \neq 0$ arises. Notice that, in this case, a Josephson current can flow even if $\chi=0$: if TRS is broken, the amplitude of processes bringing Cooper pairs from right to left lead are not necessarily compensated by those related to the opposite direction.\\
\begin{figure}
\includegraphics[width=\columnwidth,clip]{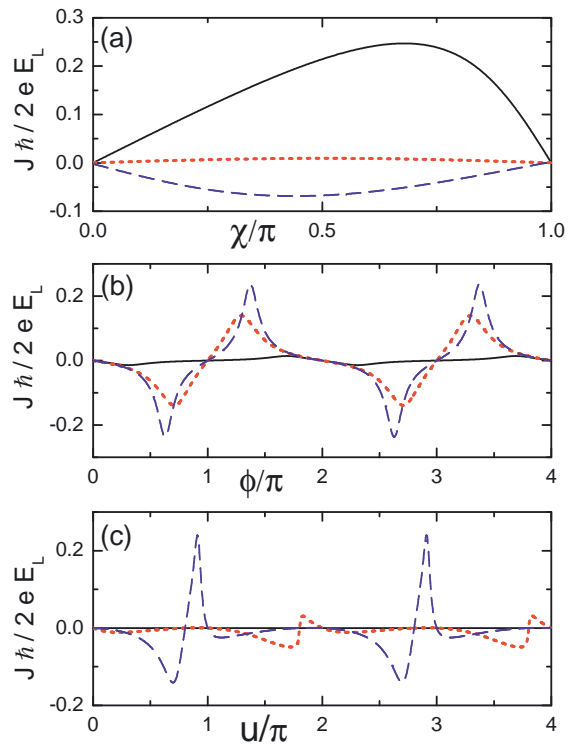}
\caption{\label{Fig2} (color online) Sign switching of the Josephson current $J$   for the set-up of Fig. \ref{AB-setup} with $|\Delta|=2 E_L$, $k_F L =60$ and contact transmission $1/2$ ($\gamma=\pi/3$).
a) $j(\chi)$  for $\phi=0$ and $u=0$ (solid), for $\phi=0$ and $u=0.8 \pi$ (dotted), and for $\phi=0.8 \pi$ and $u=0.8 \pi$ (dashed); b) $j(\chi=0)$ {\rm vs.} $\phi$ for different values of the gate voltage: $u=\pi / 20$ (dotted), $u=\pi / 2$ (dashed), and $u=1.2 \pi$ (solid); c) $j(\chi=0)$ {\rm vs.} $u$ for different values of magnetic flux: $\phi=0$ (solid), $\phi=0.1 \pi$ (dotted) and $\phi=0.9 \pi$ (dashed).}
\end{figure}

\begin{figure}
\includegraphics[width=\columnwidth,clip]{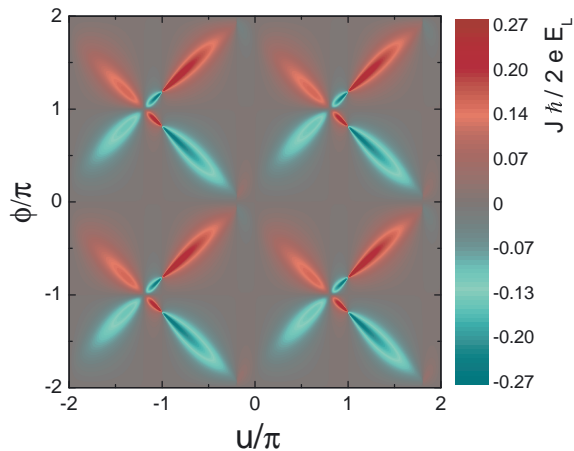}
\caption{\label{Fig3} Dimensionless Josephson current {\rm vs.} $u$ and $\phi$ for an AB ring with $|\Delta|=2 E_L$, $k_F L =60$,  $\gamma=\pi/3$, and $\chi=0$.}
\end{figure}

By substituting the $\mathsf{S}$-matrix into Eq.~(\ref{Dret-fin}), and by performing the analytic continuation $D_{\rm R}$ as discussed above, the Josephson current is determined by Eq.~(\ref{j-new}). We shall now discuss the results. Figure~\ref{Fig2} displays the switching of  $J(\chi)$ from $0$ to $\pi$-junction behavior, for a set-up with $|\Delta|=2 E_L$ and contact transmission~$1/2$ ($\gamma=\pi/3$) which corresponds, e.g., to a {\rm InGaAs} ring with $L \sim 600 \,{\rm nm}$~\cite{carillo} contacted to ${\rm Nb}$ leads, at zero temperature~\cite{NOTA-temp}. In particular, Fig.~\ref{Fig2}(a) shows $J$ as a function of the phase difference~$\chi$: when $\phi=0$ and $u=0$, the junction is in a $0$-state; by applying a gate voltage~$u=0.8 \pi$ the current is strongly suppressed (dotted curve), although its sign is always positive; however, when a magnetic flux $\phi=0.8 \pi$ is also introduced, the current sign is reversed (dashed curve). Figure~\ref{Fig2}(b) and Fig.~\ref{Fig2}(c) refer to the Josephson current for $\chi=0$, and show that its sign can be reversed by tuning either the magnetic flux~(b) or the gate voltage~(c). Notably, the curves of Fig. \ref{Fig2}(b) demonstrate the crucial role of the gate voltage: for $u=0$, although $J$ switches its sign upon varying the flux, the amplitude of the current is very small (solid curve); the latter is enhanced by increasing $u$ (dotted and dashed curves)~\cite{NOTA-curr}. The curves of Fig. \ref{Fig2}(c) refer to different values of magnetic flux; in particular, when no magnetic flux is present (solid curve) the TRS yields $J(\chi=0) \equiv 0$ independent of the gate voltage $u$, whereas when $\phi \neq 0$ a supercurrent can flow and its sign varies with~$u$ (dotted curve), with a strong enhancement when the flux approaches $\phi=\pi$.
The whole behavior of the Josephson current $J(\chi=0)$ as a function of $u$ and $\phi$ is displayed in Fig.~\ref{Fig3}. The supercurrent exhibits lobes of alternate signs whose nodes are located around some special values of $\phi$  and $u$. The latter can be roughly estimated through the condition $\cos\chi_0 \simeq -1$  (see Eq.(\ref{coschi0})), yielding $\phi \simeq (2 m_1+1) \pi$, $u \simeq (2 m_2+1) \pi$ and $u \simeq 2 ( \pi m_3-k_F L)$, with integer~$m_j$'s~\cite{NOTA}. The sign reversal is thus easily controlled around these values. This demonstrates the full tunability of the supercurrent  through electro-magnetostatic AB effect.  

In conclusion, we have proposed Aharonov-Bohm interferometry as a novel method to realize a controllable Josephson $\pi$-junction.  While the magnitude of the supercurrent can be tuned by either the electrostatic or the magnetic AB effect, its sign can be controlled only if time-reversal symmetry is broken. We have also shown that a magnetic field alone does not provide an efficient tuning of the supercurrent (see Fig.~\ref{Fig2}(b)). By contrast, the combined use of  magnetic   and electric field enhances the supercurrent amplitude, allowing a full control of the junction.   In addition, our results also imply that {\it supercurrent} measurements can be used to determine the asymmetry of the $\mathsf{S}$-matrix;   indeed  the ordinary  DC current in an   ring   contacted to normal leads only depends on the transmission coefficient $\mathcal{T}=|\tau|^2=|\tau^*|^2$, and the asymmetry cannot be probed.
\\We acknowledge stimulating discussions with R.~Fazio, F.~Carillo, A.~Khaetskii, M.~Governale, H.~Grabert and A.~Saracco, and partial financial support by HYSWITCH EU Project and MIUR   FIRB Pr. No. RBNE01FSWY.


\begin{thebibliography}{99}
\bibitem{NOTA-temp} We expect our results to be robust to thermal fluctuations for  $T < T_0 = E_L/k_B \sim  \, 10$~K.
\end{thebibliography}
\end{document}